\documentclass[aps,prd,amsmath,amssymb,showpacs]{revtex4}
\usepackage{graphicx}
\begin{document}

\title{$ZZ\gamma$ and $Z\gamma\gamma$ couplings at linear 
$e^{+} e^{-}$ collider energies  with the effects of 
Z polarization and initial state radiation }

\author{S. Ata\u{g} }
\email[]{atag@science.ankara.edu.tr}
\affiliation{Department of Physics, Faculty of Sciences, 
Ankara University, 06100 Tandogan, Ankara, Turkey}

\author{\.{I}. \c{S}ahin}
\email[]{sahin@science.ankara.edu.tr}
\affiliation{Department of Physics, Faculty of Sciences,
Ankara University, 06100 Tandogan, Ankara, Turkey}

\begin{abstract}
The constraints on the neutral gauge boson couplings, 
$ZZ\gamma$ and $Z\gamma\gamma$, are investigated 
at linear $e^{+}e^{-}$ collider energies  
through the $Z\gamma$  production 
with longitudinal and transverse  
polarization states of the final Z boson. 
Because of  high energy of linear electron-
positron beams, initial state radiation (ISR) considerably 
changes the production cross section. We obtain an increase 
in the cross section by a factor of 2-3 due to ISR
for transverse polarization and by a factor of 10-100 for
longitudinal polarization states depending 
on the energy.

We find the  95\% C.L.
limits on the CP conserving form factors  $h_{3}^{Z}$, $h_{4}^{Z}$, 
$h_{3}^{\gamma}$ and $h_{4}^{\gamma}$
with integrated luminosity 500$fb^{-1}$ and  
$\sqrt{s}=$0.5, 1, 1.5  TeV energies. 
It is shown that the longitudinal polarization of the Z 
boson, together with ISR, can improve sensitivities 
by  factors 2-3.  
\end{abstract}


\maketitle

\section{Introduction}
The self interactions of the gauge bosons are consequences 
of the non-abelian structure of the electroweak sector of the 
Standard Model (SM). 
The study of the trilinear gauge boson couplings leads to  
important tests of electroweak interactions. 
Neutral gauge boson couplings $ZZ\gamma$ and $Z\gamma\gamma$ 
are not generated at tree level by the SM. 
Higher order loop level corrections are higly below the 
current experimental sensitivity \cite{renard}.
Nevertheless, the new physics with energy scale above
present experimental threshold might provide tree level 
neutral triple-gauge boson couplings.
Deviation of the couplings from the expected values 
would indicate the existence of new physics beyond the SM. 
Therefore precision measurements of triple vector 
boson vertices will be the crucial tests of the 
structure of the SM.

For the process $e^{+}e^{-}\to Z\gamma$
it is convenient to study 
the anomalous  neutral triple gauge couplings
$Z\gamma Z^{\star}$ and $Z\gamma\gamma^{\star}$
which obey Lorentz and gauge invariance.
Within the formalism of Ref. \cite{berger,renard} there
are eight anomalous coupling parameters
$h_{i}^{Z}$, $h_{i}^{\gamma}$ $(i=1,..,4)$
which are all zero in the standard model.
Here we are interested in CP-even couplings
that are proportional to $h_{3}^{V}$ and $h_{4}^{V}$
($V=Z, \gamma$).
The photon and the Z boson in the final state are considered 
as on-shell particles while the third boson at the vertex, 
the s-channel internal propagator, is of-shell. 
Due to partial wave unitarity constraints
at high energies, an energy dependent
form factor ansatz can be cosidered:
\begin{eqnarray}
h_{i}^{V}(s)=
\frac{h_{i0}^{V}}{(1+s/\Lambda^{2})^{3}}
~~;~~i=1,3 \\
h_{i}^{V}(s)=
\frac{h_{i0}^{V}}{(1+s/\Lambda^{2})^{4}}
~~;~~i=2,4
\end{eqnarray}
In this work we assume that new physics scale $\Lambda$ is above
the collision energy $\sqrt{s}$ and we neglect the energy 
dependence of the form factors in the energy region we are 
intereted in.

CP conserving anomalous
$Z(p_{1})\gamma(p_{2})Z(p_{3})$
vertex function  can be written following the low energy 
parametrization of the residual effect from the effective 
lagrangian \cite{berger,renard}:

\begin{eqnarray}
ig_{e}\Gamma_{Z\gamma Z}^{\alpha\beta\mu}(p_{1},p_{2},p_{3})
=ig_{e}\frac{p_{3}^{2}-p_{1}^{2}}{m_{Z}^{2}} \left[h_{3}^{Z}
\epsilon^{\mu\alpha\beta\rho}p_{2\rho}+\frac{h_{4}^{Z}}{m_{Z}^{2}}
p_{3}^{\alpha}\epsilon^{\mu\beta\rho\sigma}p_{3\rho}p_{2\sigma}
\right]
\end{eqnarray}
where $m_{Z}$ and $g_{e}$ are the Z-boson mass and charge of the
proton.
The $Z\gamma\gamma$ vertex function
can be obtained with the replacements:

\begin{eqnarray}
\frac{p_{3}^{2}-p_{1}^{2}}{m_{Z}^{2}}\to
\frac{p_{3}^{2}}{m_{Z}^{2}} ,\;\;\; h_{i}^{Z}\to h_{i}^{\gamma}
,\;\;\; i=3,4
\end{eqnarray}
The overall factor $p_{3}^{2}$ in the $Z\gamma\gamma^{\star}$
vertex function originates from electromagnetic gauge invariance.
Due to Bose statistics  the $Z\gamma\gamma$ 
vertex vanishes identically if both
photons are on shell (Yang's theorem) \cite{yang}.

Previous limits on the $Z\gamma Z^{\star}$ 
and $Z\gamma\gamma^{\star}$
anomalous couplings have been provided by the Tevatron 
$|h_{3}^{Z}|<0.36$, $|h_{4}^{Z}|<0.05$, 
$|h_{3}^{\gamma}|<0.37$ and $|h_{4}^{\gamma}|<0.05$
\cite{d0} and the combination of four LEP 
experiments ALEPH, DELPHI, L3, OPAL $-0.20<h_{3}^{Z}<0.07$, 
$-0.05<h_{4}^{Z}<0.12$,
$-0.049<h_{3}^{\gamma}<0.008$ and 
$-0.002<h_{4}^{\gamma}<0.034$
\cite{l3} at 95\% C.L. .
Based on the analysis of ZZ production at the 
upgraded Fermilab Tevatron and the CERN Large Hadron 
Collider (LHC) achievable limits on the
 $ZZ\gamma$ couplings 
( $f_{4}^{Z}$,  $f_{4}^{\gamma}$, 
$f_{5}^{Z}$,  $f_{5}^{\gamma}$)
have been discussed \cite{baur,renard2}.

Research and development on linear $e^{+}e^{-}$
colliders at SLAC, DESY and KEK have been progressing
and the physics potential of these future machines
is under intensive study.
In this paper, the process 
$e^{+}e^{-}\to Z\gamma $ with the final state
$ \ell^{+}\ell^{-} \gamma $ 
is investigated to search for 
$Z\gamma Z^{\star}$ and $Z\gamma\gamma^{\star}$
couplings. Because of the high energy of the incomig beams, 
initial state electromagnetic radiative 
corrections (ISR) are taken into account using the structure 
function method.
Another important point is the polarization of the final 
state Z boson. In order to determine the Z polarization,
we show that the angular distribution of 
the Z decay products,  has a clear correlation 
with the helicity states of the Z boson. 

\section{Cross Sections and Angular Correlations for 
Final State Fermions}

In this section, we present the cross section calculation  
via helicity amplitudes for the complete process 
$e^{+}e^{-}\to Z\gamma \to \ell^{+}\ell^{-} \gamma$
and describe angular distributions of final state fermions 
to see the correlations with the polarization states of Z boson.
Let us start with the differential cross section 

\begin{eqnarray}
d\sigma(e^{+}e^{-}\to Z\gamma \to \ell^{+}\ell^{-} \gamma)=
\frac{1}{2s}|M|^{2}\frac{d^{3}p_{3}}{(2\pi)^{3}2E_{3}} 
\frac{d^{3}p_{4}}{(2\pi)^{3}2E_{4}}
\frac{d^{3}k_{2}}{(2\pi)^{3}2E_{\gamma}}
(2\pi)^{4}\delta^{4}(p_{1}+p_{2}-p_{3}-p_{4}-k_{2})
\end{eqnarray}
where $p_{1}, p_{2}$ are  the momenta of incoming leptons,
 $p_{3}, p_{4}$ are the momenta of outgoing fermions and
$k_{2}$ is the momentum of outgoing  photon. $|M|^{2}$ is 
the square of the full  amplitude which is averaged over 
initial spins and summed over final spins. The helicity 
dependent full amplitude  can be expressed as follows 

\begin{eqnarray}
|M(\sigma_{1},\sigma_{2}; \sigma_{3},\sigma_{4}, 
\lambda_{\gamma})|^{2}
(2\pi)^{4}\delta^{4}(p_{1}+p_{2}-p_{3}-p_{4}-k_{2})&&=
\int{\frac{d^{4}k_{1}}{(2\pi)^{4}}} |\sum_{\lambda_{Z}}
M_{a}(\sigma_{1},\sigma_{2};\lambda_{Z},\lambda_{\gamma})
 D_{Z}(k_{1}^{2}) M_{b}(\lambda_{Z};\sigma_{3},\sigma_{4}|^{2}
\nonumber \\
&&(2\pi)^{4} \delta^{4}(p_{1}+p_{2}-k_{1}-k_{2})
(2\pi)^{4} \delta^{4}(k_{1}-p_{3}-p_{4})
\end{eqnarray}
where $k_{1}$ is the internal momentum of the Z boson propagator.
$\sigma_{1}, \sigma_{2} $ are the incoming lepton helicities,
$\sigma_{3},\sigma_{4}, \lambda_{\gamma}$ are the outgoing fermion 
and photon helicities. For the summation over the intermediate 
Z boson polarization we take the helicity basis $\lambda_{Z}=+,-,0$. 
Here $D_{Z}(k_{1})$ is the Breit-Wigner propagator factor for the 
Z boson 
\begin{eqnarray}
D_{Z}(k_{1})=\frac{1}{k_{1}^{2}-M_{Z}^{2}+iM_{Z}\Gamma_{Z}}
\end{eqnarray}
Here $M_{a}(\sigma_{1},\sigma_{2};\lambda_{Z},\lambda_{\gamma})$
is the helicity amplitudes for $e^{+}e^{-}\to Z\gamma $ with on shell 
Z boson which are provided in the appendix. 
$M_{b}(\lambda_{Z};\sigma_{3},\sigma_{4})$ is the decay 
amplitude of the Z boson which will be given later.
Using the narrow width approximation differential cross 
section becomes 
\begin{eqnarray}
d\sigma(e^{+}e^{-}\to Z\gamma \to \ell^{+}\ell^{-} \gamma)&&=
\frac{1}{2s}\frac{1}{(2M_{Z}\Gamma_{Z})}
|\sum_{\lambda_{Z}} M_{a}(\lambda_{Z})M_{b}(\lambda_{Z})|^{2}
\nonumber \\
&&\frac{d^{3}k_{1}}{(2\pi)^{3}2E_{Z}}
\frac{d^{3}k_{2}}{(2\pi)^{3}2E_{\gamma}}
(2\pi)^{4} \delta^{4}(p_{1}+p_{2}-k_{1}-k_{2})
\nonumber \\
&&\frac{d^{3}p_{3}}{(2\pi)^{3}2E_{3}}
\frac{d^{3}p_{4}}{(2\pi)^{3}2E_{4}}
(2\pi)^{4} \delta^{4}(k_{1}-p_{3}-p_{4})
\end{eqnarray}
Here $M_{a}(\lambda_{Z})$ and $M_{b}(\lambda_{Z})$ indicate 
average over initial lepton spins, summmation over final state 
fermion spins and photon.
In the rest frame of the Z boson where the decay amplitudes 
are most simply expressed, we get  

\begin{eqnarray}
d\sigma(e^{+}e^{-}\to Z\gamma \to \ell^{+}\ell^{-} \gamma)&&=
\frac{1}{2s}\frac{1}{((2\pi)^{2}16M_{Z}\Gamma_{Z})}
|\sum_{\lambda_{Z}} M_{a}(\lambda_{Z})M_{b}(\lambda_{Z})|^{2}
\nonumber \\
&&\frac{d^{3}k_{1}}{(2\pi)^{3}2E_{Z}}
\frac{d^{3}k_{2}}{(2\pi)^{3}2E_{\gamma}}
(2\pi)^{4} \delta^{4}(p_{1}+p_{2}-k_{1}-k_{2})
\nonumber \\
&&d\cos{\theta^{\star}} d\phi^{\star}
\end{eqnarray}
where $\theta^{\star}$ and $\phi^{\star}$ are the polar and 
azimuthal angles of the final state leptons in the Z rest frame 
with respect to the Z boson direction in the 
$\ell\ell\gamma$ rest frame.
After integration over azimuthal angles $\phi^{\star}$ 
interference terms will vanis as below

\begin{eqnarray}
d\sigma(e^{+}e^{-}\to Z\gamma \to \ell^{+}\ell^{-} \gamma)&&=
\frac{1}{2s}\frac{1}{(32\pi M_{Z}\Gamma_{Z})}
[|M_{a}(+)M_{b}(+)|^{2}+|M_{a}(-)M_{b}(-)|^{2}+
|M_{a}(0)M_{b}(0)|^{2}]
\nonumber \\
&&\frac{d^{3}k_{1}}{(2\pi)^{3}2E_{Z}}
\frac{d^{3}k_{2}}{(2\pi)^{3}2E_{\gamma}}
(2\pi)^{4} \delta^{4}(p_{1}+p_{2}-k_{1}-k_{2})
d\cos{\theta^{\star}} 
\end{eqnarray}
This can be written in a more compact form

\begin{eqnarray}
d\sigma(e^{+}e^{-}\to Z\gamma \to \ell^{+}\ell^{-} \gamma)&&=
\frac{1}{(32\pi M_{Z}\Gamma_{Z})}
\left[|d\sigma_{1}(+)|M_{b}(+)|^{2}+d\sigma_{1}(-)|M_{b}(-)|^{2}+
d\sigma_{1}(0)|M_{b}(0)|^{2}\right]
d\cos{\theta^{\star}}
\end{eqnarray}

If one performs further integration over polar angle 
$\cos{\theta^{\star}}$ it turns out to be a well known result 

\begin{eqnarray}
d\sigma(e^{+}e^{-}\to Z\gamma \to \ell^{+}\ell^{-} \gamma)&&=
[d\sigma_{1}(+)+d\sigma_{1}(-)+d\sigma_{1}(0)] 
BR(Z\to \ell^{+}\ell^{-}) \nonumber \\
&&=d\sigma_{1}(e^{+}e^{-}\to Z\gamma)BR(Z\to \ell^{+}\ell^{-})
\end{eqnarray}

\begin{figure}
\includegraphics{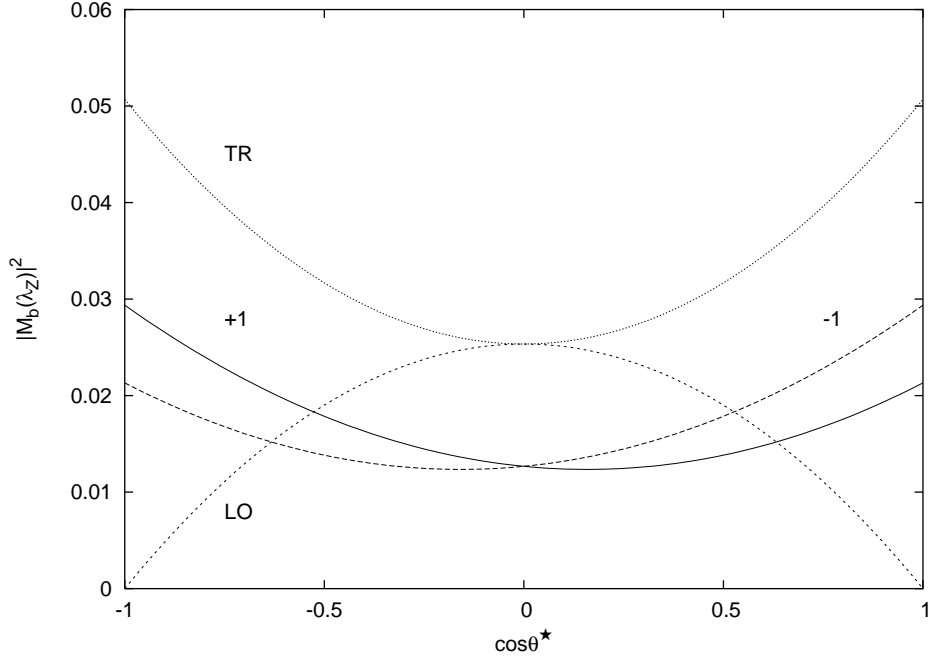}
\caption{Polar angle distributions of Z decay product 
$(1/32\pi\Gamma_{Z})|M_{b}(\lambda_{Z})|^{2}$ in the rest 
frame of Z boson  for $\lambda_{Z}=-1,+1,0$. Transverse 
polarization state is defined as the sum of +1 and -1 states.
\label{fig1}}
\end{figure}

Explicit forms of the decay amplitudes $|M_{b}(+)|^{2}$, 
$|M_{b}(-)|^{2}$ and $|M_{b}(0)|^{2}$ in the Z rest frame are 
given by

\begin{eqnarray}
|M_{b}(+)|^{2}&&=\frac{M_{Z}^{2}}{2}
[g_{L}^{2}(1-\cos{\theta^{\star}})^{2}+
g_{R}^{2}(1+\cos{\theta^{\star}})^{2}]
\\ 
|M_{b}(-)|^{2}&&=\frac{M_{Z}^{2}}{2}
[g_{L}^{2}(1+\cos{\theta^{\star}})^{2}+
g_{R}^{2}(1-\cos{\theta^{\star}})^{2}]
\\
|M_{b}(0)|^{2}&&=M_{Z}^{2}(g_{L}^{2}+g_{R}^{2})
\sin^{2}{\theta^{\star}}
\\
|M_{b}(TR)|^{2}&&=|M_{b}(+)|^{2}+|M_{b}(-)|^{2}
\\
|M_{b}(LO)|^{2}&&=|M_{b}(0)|^{2}
\end{eqnarray}
By measuring the polar angle distributions of the 
Z decay products one can directly determine 
the differential cross sections for 
fixed Z helicities. In other words,  
Z helicity states are obtained from a fit to 
these distributions. Complete factors 
$(1/32\pi\Gamma_{Z})|M_{b}(\lambda_{Z})|^{2}$
in front of $d\sigma_{1}(\lambda_{Z})$ in differential 
cross sections are plotted in Fig.\ref{fig1}.
As can be seen from Fig.\ref{fig1} longitudinal(LO) and 
transverse (TR) distributions are well separated from 
each other. This is why we consider only transverse and 
longitudinal polarizations for the Z boson in next sections.

\section{Initial State Electromagnetic Radiative Correction}
Due to small mass of the electron, a significant role 
is played by the electromagnetic radiative corrections 
to the initial electron-positron state epecially at 
linear collider  energies.
In this work we use structure function formalism to describe 
the electromagnetic radiative corrections in $e^{+}e^{-}$ 
colliders \cite{fadin} .
The cross section can be written in the following form within 
this formalism

\begin{eqnarray}
\sigma(s)=\int{dx_{1}}\int{dx_{2}}~ D_{1}(x_{1},Q^{2})~ 
D_{2}(x_{2},Q^{2}) \sigma^{\prime}(s^{\prime})
\end{eqnarray}
where $\sigma^{\prime}(s^{\prime})$ is the cross section 
with reduced energy $s^{\prime}=x_{1}x_{2}s$.
$D_{1}(x_{1},Q^{2})$ ($D_{2}(x_{2},Q^{2})$) stands for the electron 
(positron) structure function giving the probability of finding 
an electron (positron) within an electron (positron) with
a longitudinal momentum fraction $x_{1}$ ($x_{2}$). Although 
several definitions of the structute functions are present
we use the following ones which are used by HERWIG 
\cite{herwig}

\begin{eqnarray}
D_{1}(x,s)&&=\beta (1-x)^{\beta-1} g(x,s)
\\
g(x,s)&&=e^{\beta (1+x/2)x/2} (1-\beta^{2} \frac{\pi^{2}}
{12})+y\frac{\beta^{2}}{8} y [(1+x)\{(1+x)^{2}+3 \log{x}\}
-\frac{4 \log{x}}{1-x}]
\\
\beta&&=\frac{\alpha_{em}}{\pi}(\log{\frac{s}{M_{e}^{2}}-1})~,~~~
y=[\beta (1-x)^{\beta-1}]^{-1}
\end{eqnarray}

To avoid divergency at the upper limit of the momentum fraction,
$x=1$, the cross section can be transformed into different form

\begin{eqnarray}
\sigma(s)&&=\int_{0}^{1-\epsilon}{dx_{1}}
\int_{0}^{1-\epsilon}{dx_{2}}~ 
D_{1}(x_{1},s)~
D_{2}(x_{2},s) \sigma^{\prime}(s^{\prime}) \nonumber \\
&&+\epsilon^{\beta} g_{2}(1,s) \int_{0}^{1-\epsilon}{dx_{2}}~
D_{1}(x_{1},s)\sigma^{\prime}(x_{1}s) +
\epsilon^{\beta} g_{1}(1,s) \int_{0}^{1-\epsilon}{dx_{2}}~
D_{2}(x_{2},s)\sigma^{\prime}(x_{2}s)\nonumber \\
&&+\epsilon^{2\beta} g_{1}(1,s) g_{2}(1,s) \sigma^{\prime}(s)
\end{eqnarray}
where $\epsilon$ can be taken as $10^{-9}-10^{-12}$. In this 
region of $\epsilon$ the cross section changes by a factor of 0.985. 
If one takes smaller $\epsilon$ values,  higher machine precision 
gives softer $\epsilon$ dependence. 
The following transformation gives relatively smooth integrand

\begin{eqnarray}
\int_{0}^{1-\epsilon}{dx_{1}}
\int_{0}^{1-\epsilon}{dx_{2}}~
D_{1}(x_{1},s)~
D_{2}(x_{2},s) \sigma^{\prime}(s^{\prime})=
\int_{E_{min}}^{E_{max}}{dE}
\int_{F_{min}}^{F_{max}}{dF}~
g_{1}(x_{1},s)~
g_{2}(x_{2},s) \sigma^{\prime}(x_{1}x_{2}s)
\end{eqnarray}
where

\begin{eqnarray}
x_{1}&&=1-(-F)^{1/\beta}~,~~~~x_{2}=1-(-E)^{1/\beta}
\\
E_{min}&&=-(1-\frac{\tau_{min}}{1-\epsilon})~,
~~~E_{max}=-\epsilon^{\beta} \\
F_{min}&&=-(1-\frac{\tau_{min}}{x_{2}})~,
~~~F_{max}=-\epsilon^{\beta} \\
\tau_{min}&&=\frac{M_{Z}^{2}}{s}
\end{eqnarray}

\begin{figure}
\includegraphics{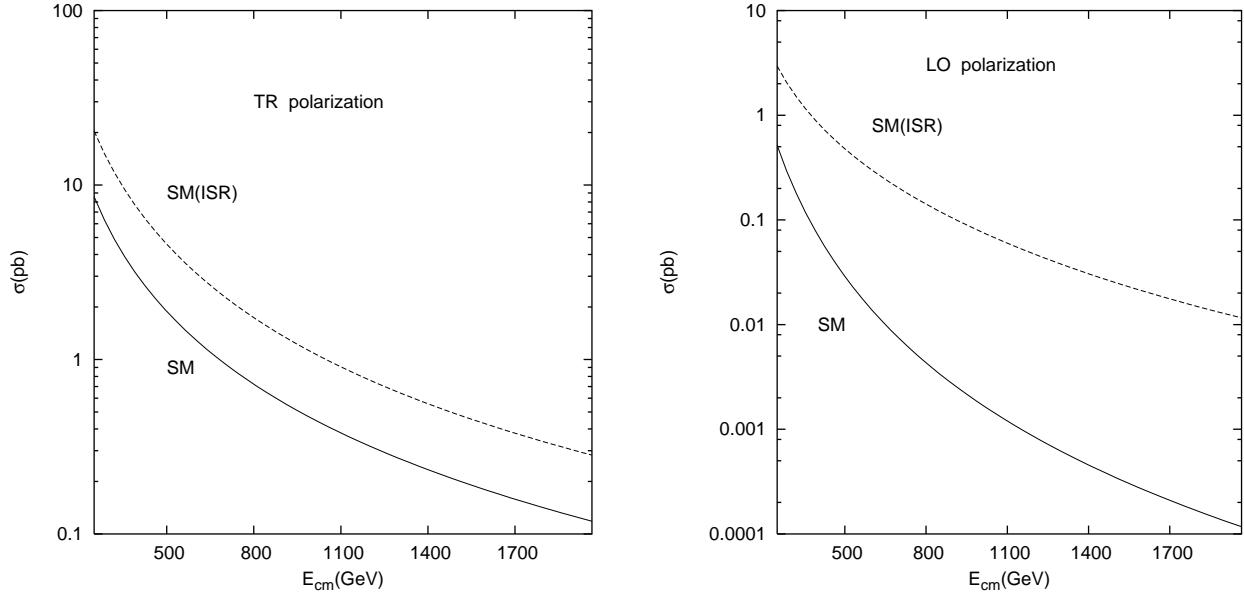}
\caption{Effect of ISR to the standard model cross section 
for $e^{+}e^{-}\to \gamma Z$ with transverse and 
longitudinal polarization of Z boson. Two polarization 
states, TR and LO, get different contributions from ISR.
\label{fig2}}
\end{figure}

Using above formalism we calculate the cross section with
initial state radiative corrections for both Standard Model 
and anomalous coupling cases.
In Fig.\ref{fig2} the effects of ISR to 
the total cross section for the standard model process
$e^{+}e^{-}\to \gamma Z$ 
are shown as a function of energy $\sqrt{s}$ (before ISR) 
with transverse(TR) and longitudinal(LO) Z polarization.
Here average over initial spins and sum over photon polarization
are performed. The unpolarized cross section  is 
almost the same as the TR polarization case since the magnitude 
of the cross secion is dominated by the cross section with 
TR polarization. 
As can be seen from the figures, ISR increases the 
cross section with a factor 2-3 for the TR case and 
a factor of 10-100 for LO polarization case depending on the 
energy. Furhtermore, very small LO cross section 
becomes sizable due to ISR. The reason for this  
comes from the fact that the energy dependences of the 
LO and TR cross sections are different.
In order to understand this feature let us write standard model 
squared amplitudes for transverse and longitudinal Z 
polarizations using the helicity amplitudes given in the 
appendix

\begin{eqnarray}
|M_{a}(TR)|^{2}=&&8[(C_{3}^{L})^{2}+(C_{3}^{R})^{2}]
(\frac{2}{\sin^{2}{\theta}}-1)
\frac{(m_{Z}^{4}+s^{2})}{(s-m_{Z}^{2})^{2}} \\
|M_{a}(LO)|^{2}=&&16[(C_{3}^{L})^{2}+(C_{3}^{R})^{2}]
\frac{m_{Z}^{2}s}{(s-m_{Z}^{2})^{2}}
\end{eqnarray}
where  $M_{a}(\lambda_{Z})$ is defined in Eq.(8) 
and we neglect electron mass.
As can be seen here the longitudinal part is independent 
of the polar angle  whereas the transverse part strongly depends 
on it. Angular integration implies that contribution of 
the transverse polarization always gives larger cross 
section than the longitudinal one. 
In both cases, the largest cross section comes 
from the energy region where $\sqrt{s}\simeq m_{Z}$.
When we consider ISR, the major contribution 
from the integration over 
$x_{1}$ and $x_{2}$ to the cross section is due to 
the lower limit of the 
$s^{\prime}=x_{1}x_{2}s\simeq m_{Z}^{2}$.
In this limit the above amplitudes take the forms

\begin{eqnarray}
|M_{a}(TR)|^{2}\simeq&&16[(C_{3}^{L})^{2}+(C_{3}^{R})^{2}]
(\frac{2}{\sin^{2}{\theta}}-1)
\frac{m_{Z}^{4}}{(\Delta s)^{2}} \\
|M_{a}(LO)|^{2}\simeq&&16[(C_{3}^{L})^{2}+(C_{3}^{R})^{2}]
\frac{m_{Z}^{4}}{(\Delta s)^{2}}
\end{eqnarray}
where we use $s=m_{Z}^{2}+\Delta s$ with $\Delta s<<m_{Z}^{2}$.

The cross section without ISR leads to  
$s^{\prime}=s$ with $x_{1}=x_{2}=1$.  For the collider  
energy $\sqrt{s}=1$ TeV where $s>>m_{Z}^{2}$ the amplitudes 
becomes 

\begin{eqnarray}
|M_{a}(TR)|^{2}\simeq&&8[(C_{3}^{L})^{2}+(C_{3}^{R})^{2}]
(\frac{2}{\sin^{2}{\theta}}-1)
(1+2\frac{m_{Z}^{2}}{s}) \\
|M_{a}(LO)|^{2}\simeq&&16[(C_{3}^{L})^{2}+(C_{3}^{R})^{2}]
\frac{m_{Z}^{2}}{s}
\end{eqnarray}
Here the LO cross section continiously decreases as 
$s$ increases. This is the expected result for longitudinal 
polarization. Because there is no LO polarization for masless vector 
bosons.
For $\sqrt{s}=10 m_{Z}$ the effect of 
ISR on both amplitudes can be  compared easily  as below:
\begin{eqnarray}
\frac{|M_{a}(TR)|_{ISR}^{2}} {|M_{a}(TR)|^{2}}\simeq
\frac{2m_{Z}^{4}}{(\Delta s)^{2}} \\
\frac{|M_{a}(LO)|_{ISR}^{2}} {|M_{a}(LO)|^{2}}\simeq
50\frac{2m_{Z}^{4}}{(\Delta s)^{2}} 
\end{eqnarray}

\begin{figure}
\includegraphics{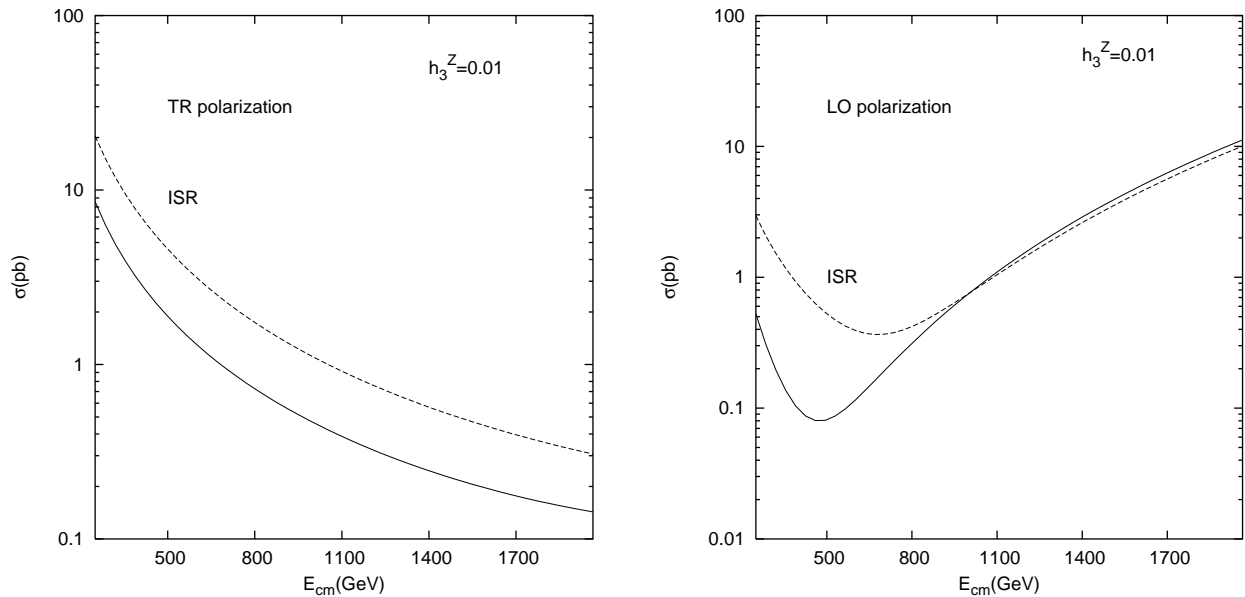}
\caption{Energy dependence of the total cross section for
$e^{+}e^{-}\to \gamma Z$ with anomalous $ZZ\gamma$ vertex
parameter $h_{3}^{Z}=0.01$. Effects of ISR and Z polarization
are also shown. As in the previous figure, 
two polarization states of the 
Z boson  get different contributions from ISR.
\label{fig3}}
\end{figure}

\begin{figure}
\includegraphics{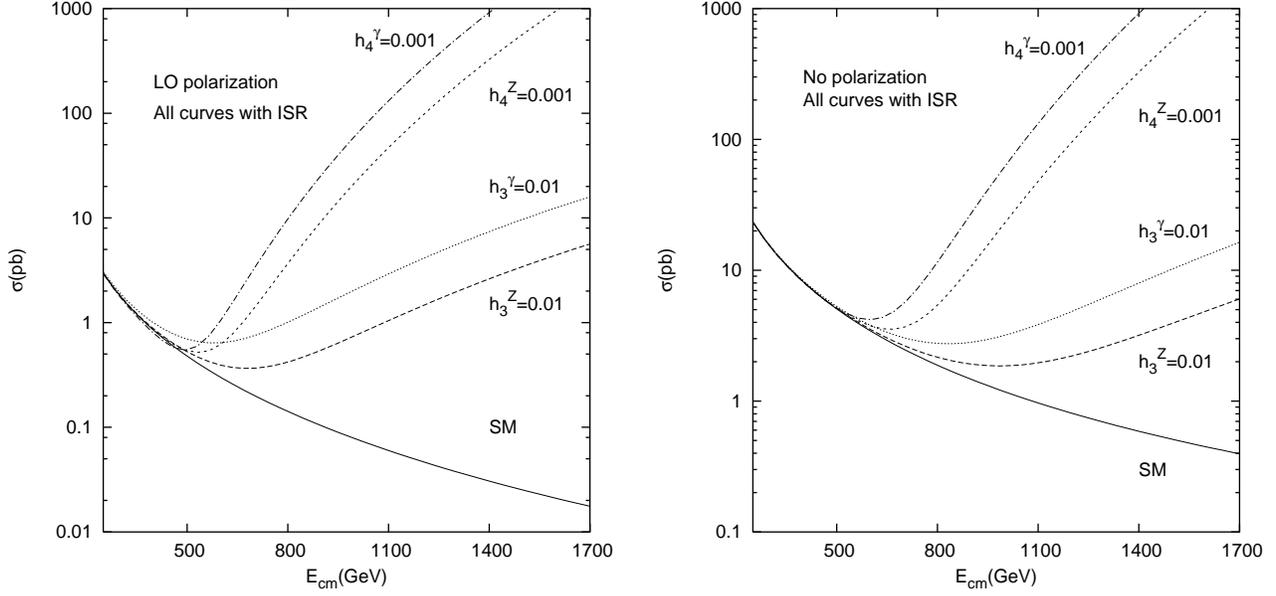}
\caption{ Energy dependence of the total cross section for
$e^{+}e^{-}\to \gamma Z$ with LO polarization (left) and 
unpolarized (right) cases. The values of the anomalous 
$ZZ\gamma$ and $Z\gamma\gamma$ vertex parameters 
$h_{3}^{Z}=h_{3}^{\gamma}=0.01$ and 
$h_{4}^{Z}=h_{4}^{\gamma}=0.001$ are chosen.
All cross sections are ISR corrected. 
\label{fig4}}
\end{figure}

Fig.\ref{fig3} shows the energy dependence of the total 
cross sections with anomalous coupling parameter $h_{3}^{Z}=0.01$
for TR and LO polarization states. The contributions of
anomalous couplings become remarkably important after the 
center of mass energy 500-600 GeV.
Effect of ISR is also shown for two cases. 
In the case of LO polarization, after the 
energy of 900 GeV the ISR  gives negligible corrections.
Similar behaviour appears for $h_{3}^{\gamma}=0.01$
and $h_{4}^{Z}=h_{4}^{\gamma}=0.001$ values.
In Fig.\ref{fig4} the energy dependences of the ISR corrected 
total cross sections with four anomalous coupling parameters 
are plotted for LO polarization and unpolarized cases. 

\begin{figure}
\includegraphics{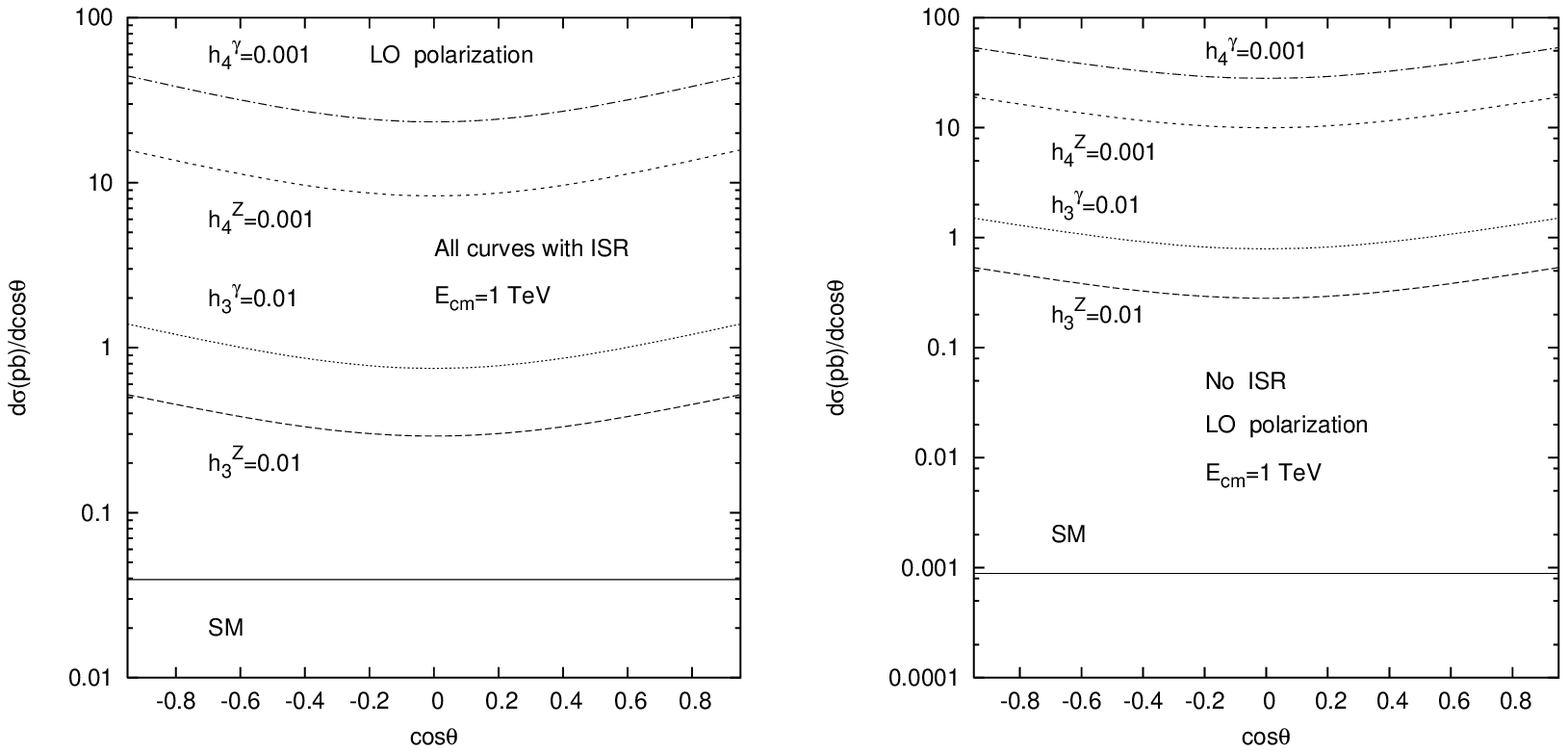}
\caption{Angular distributions of the outgoing Z boson
in the center of mass frame of outgoing Z and the photon.
$\theta$ is the angle between the outgoing Z boson and the 
incoming electron. Angular dependence of the 
four anomalous couplings are shown 
with (left) and without(right) ISR corrections.
Only LO polarization states are taken into account in these 
figures since TR polarization states are  poorly sensitive 
to anomalous couplings.
\label{fig5}}
\end{figure}

It is also important to see how the anomalous couplings 
change the shape of the angular distribution of the
Z boson for the polarized and unpolarized cases. Angular 
distribution of the Z boson is shown in Fig. \ref{fig5}
for LO polarization state with and wihout ISR correction.
Since the TR polarization state 
of the Z boson is poorly sensitive to anomalous couplings, 
angular distributions with TR case are not plotted.
In all figures,  only one of the coupling
parameters are kept different from zero.
From all these figures we reach at the following 
remarkable results.  
The LO polarization states 
are always more sensitive to the anomalous couplings.
Much larger deviations  arise from  $h_{4}^{V}$ for
both TR and LO polarizations.
$Z\gamma\gamma^{\star}$ couplings always provide the
higher contribution to the cross section than the
$Z\gamma Z^{\star}$ couplings.
The shape of the  curves differs for two 
kinds of polarizations of the Z boson.
The ISR correction gives larger contribution to the  
SM cross section than the case 
with anomalous couplings. Therefore, the sensitivity of 
the $e^{+}e^{-}\to \gamma Z$ process to the anomalous 
couplings is expected to become poorer 
due to the ISR correction. 
Numerical results for all polarization 
configurations will be given in the next section.

\section{Limits on the Anomalous Coupling Parameters}

If the Z boson decays into a pair of charged leptons 
the signal for the final states can be $\gamma \ell \ell$
where we consider $\ell =e, \mu$. The potential 
background processes for $\ell=\mu$ final state are the following:\\ 

$e^{+}e^{-}\to Z(\gamma) \to \gamma \ell \ell$ :  
s channel Z or $\gamma$ exchange (final state bremstrahlung). \\

$e^{+}e^{-}\to \gamma\gamma \to \gamma \ell \ell$ : 
t channel $e$ exchange. \\

For $\ell=e$ final state additional t channel background 
processes such as those arising from both 
Z($\gamma$) and e exchange;  Z or $\gamma$ 
exchange (final state bremstrahlung) are present.
Since we take into account only on shell Z bosons, 
we should impose  a cut on the invariant mass
of charged leptons $M_{\ell\ell}\simeq M_{Z}$. 
This cut reduces the effect of  background processes for 
$\ell =\mu$ drastically. 
The total cross section of background processes at least 
100 times smaller than the process 
$e^{+}e^{-}\to Z \gamma \to \gamma \ell \ell$ (t channel 
$e$ exchange with on shell Z boson). 
In the case of $\ell =e$ final state
background cross sections are  10 times higher
when compared to $\mu$ final states. 
Therefore the major potential background is due 
to $\ell =e$ final states.
In the following sensitivity calculation, background 
contributions have a negligible  effect.
In order to obtain realistic limits on the $h_{3}^{V}$
and $h_{4}^{V}$ from the linear  $e^{+}e^{-}$ collider
the number of events have been calculated 
using 
$N=A \sigma(e^{+}e^{-}\to\gamma Z) BR(Z \to \ell^{+}\ell^{-} )L_{int}$ 
for integrated luminosity $L_{int}=500 fb^{-1}$.
Here lepton channel of the Z decay  and overall acceptance 
A=0.85 has been taken into account. For total cross section 
$|\cos{\theta}|=0.99$ has been used as angular region.
The 95\% confidence level (C.L.) limits have been estimated 
from total cross section 
using simple one parameter $\chi^{2}$ test for 
$\sqrt{s}=$0.5, 1, 1.5 TeV. 
At $\sqrt{s}=1.5$ TeV the SM cross section without 
ISR correction gives the smallest number of events N=40 
(using the above formula) for LO polarization state.
If ISR correction is included  number of events 
increases up to N=2800. For lower initial 
energy, $\sqrt{s}=1$ TeV, the smallest event number 
becomes  250 and 10000 without and with ISR correction. 
The limits which have been  obtained are shown in Tables 
\ref{tab1}-\ref{tab3} for the deviation of the cross section
from the standard model value without systematic error.
It should be noted that better limits are obtained for the 
polarization configuration $\lambda_{Z}=$LO
which leads to the order of O$(10^{-3}-10^{-4})$ for 
$h_{3}^{Z}$ and $h_{3}^{\gamma}$, O$(10^{-4}-10^{-6})$ for 
$h_{4}^{Z}$ and $h_{4}^{\gamma}$ at $\sqrt{s}=0.5-1.5$ TeV
when ISR corrections are taken into account.
Without ISR correction, the LO  polarization 
of the Z boson improves the limits 
by factors 3-7 depending on the energy.
The LO  polarization with ISR correction 
improves the limits by factors 2-2.5 which should 
be taken as the realistic results.
As can be seen from the tables, TR polarizations 
are not sensitive to anomalous couplings.
But the advantage of TR polarization case is the 
absence of $h_{4}^{Z}$ and $h_{4}^{\gamma}$ couplings.
This feature reduces the number of coupling parameters.
In order to see the degree of energy dependence on the 
anomalous couplings let us take into account the increase 
in c.m. energy from 0.5 TeV to 1.5 TeV for the 
LO polarization configuration. Then we get  
the improvements in  sensitivity limits by a factor 20, 30 
for $h_{3}^{Z}$ and $h_{3}^{\gamma}$, 
by a factor 150, 200 for $h_{4}^{Z}$ and 
$h_{4}^{\gamma}$.

\section{Conclusion}

Future linear $e^{+}e^{-}$ colliders with $\sqrt{s}=0.5-1.5$ TeV
energy and integrated luminosity 500$fb^{-1}$
probe the $Z\gamma Z^{\star}$ and $ Z\gamma\gamma^{\star}$
anomalous couplings with far better sensitivity than the
present colliders Fermilab Tevatron and LEP2 experiments.
Measurement of final state Z boson polarization is important 
in two ways. First, LO polarization state is always far more 
sensitive to anomalous couplings. Second, TR polarization state 
contains only $h_{3}^{Z}$ and  $h_{3}^{\gamma}$. 
Inital state electromagnetic radiative correction improves 
the number of events especially for the Standard Model processes 
at the linear collider energies. Furthermore, LO 
polarization case gets larger contribution from 
the ISR than the case of TR polarization.

Some limits on the above couplings via $e^{+}e^{-}\to Z\gamma$ 
process at linear collider energies from similar works  
are the following for comparison:  
the sensitivity (one standard deviation) 
$2\times 10^{-4}$ , $4\times 10^{-3}$, $4\times 10^{-5}$,
$3\times 10^{-4}$ for $h_{3}^{\gamma}$, $h_{3}^{Z}$,
$h_{4}^{\gamma}$, $h_{4}^{Z}$, respectively 
at 500 GeV energy and 100 $fb^{-1}$ luminosity \cite{renard2};
the sensitivity(95\% C.L.)  $O(10^{-2})$ 
for $h_{3}^{\gamma}$, $h_{3}^{Z}$, 
$O(10^{-3})$ for $h_{4}^{\gamma}$, $h_{4}^{Z}$ at 
500 GeV  and 10 $fb^{-1}$ \cite{walsh}.
Predictions from CERN LHC \cite{berger} 
give the limits $5.2\times 10^{-3}$, $3.7\times 10^{-5}$
for $h_{3}^{Z}$ and $h_{4}^{Z}$ at $2\sigma$, 10 $fb^{-1}$.
Comparison of the results in this work at $\sqrt{s}=1.5$ TeV 
with those of LHC shows that our results are improved one 
order of magnitude. For more precise results, 
further analysis needs to be
supplemented with a more detailed knowledge
of the experimental conditions.

\begin{table}
\caption{Sensitivity of the linear $e^{+} e^{-}$ collider 
to $ZZ\gamma$ and $Z\gamma\gamma$ couplings at 95\% C.L. for 
$\sqrt{s}=0.5$ TeV and $L_{int}=500$ $fb^{-1}$. Only one of the 
couplings is assumed to deviate from the SM at a time.\label{tab1}}
\begin{ruledtabular}
\begin{tabular}{ccccc}
$\lambda_{Z}$ &$h_{3}^{Z}$&
$h_{4}^{Z} $ & $h_{3}^{\gamma} $
&$h_{4}^{\gamma}$ \\
\hline
TR+LO &$|6\times 10^{-3}|$ & $|4\times 10^{-4}|$
& $|4\times 10^{-3}|$ & $|2\times 10^{-4}|$  \\
(ISR) &$|8\times 10^{-3}|$  & $|6\times 10^{-4}|$ & 
$|5|\times 10^{-3}$ & $|3\times 10^{-4}|$  \\
TR &$|3\times 10^{-2}|$ & -  &  $|2\times 10^{-2}|$ & -  \\
(ISR) & $|4\times 10^{-2}|$ & -   &  $|3\times 10^{-2}|$ & - \\
LO & $|2\times 10^{-3}|$  & $ |1\times 10^{-4}|$ & 
$|1\times 10^{-3}|$  & $|8\times 10^{-5}|$  \\
(ISR) & $|4\times 10^{-3}|$ & $|3\times 10^{-4}|$  & 
$|3\times 10^{-3}|$  & $|2\times 10^{-4}|$ \\
\end{tabular}
\end{ruledtabular}
\end{table}

\begin{table}
\caption{Sensitivity of the linear $e^{+} e^{-}$ collider
to $ZZ\gamma$ and $Z\gamma\gamma$ couplings at 95\% C.L. for
$\sqrt{s}=1$ TeV and $L_{int}=500$ $fb^{-1}$. Only one of the
couplings is assumed to deviate from the SM at a time.\label{tab2}}
\begin{ruledtabular}
\begin{tabular}{ccccc}
$\lambda_{Z}$ &$h_{3}^{Z}$&
$h_{4}^{Z} $ & $h_{3}^{\gamma} $
&$h_{4}^{\gamma}$ \\
\hline
TR+LO &$|1\times 10^{-3}|$& $ |2\times 10^{-5}|$ & 
$|6\times 10^{-4}|$ & $|1\times 10^{-5}|$  \\
(ISR) &$|1\times 10^{-3}|$  & $|2\times 10^{-5}|$ & 
$|8\times 10^{-4}|$ & $|1\times 10^{-5}|$  \\
TR &$|1\times 10^{-2}|$ & - &  $|7\times 10^{-3}|$ & -  \\
(ISR) & $|1\times 10^{-2}|$ & -   & $ |8\times 10^{-3}|$ & - \\
LO & $|3\times 10^{-4}|$  & $ |4\times 10^{-6}|$ & 
$|2\times 10^{-4}|$  & $|3\times 10^{-6}|$  \\
(ISR) & $|7\times 10^{-4}|$ & $|1\times 10^{-5}|$  & 
$|4\times 10^{-4}|$  & $|7\times 10^{-6}|$ \\
\end{tabular}
\end{ruledtabular}
\end{table}

\begin{table}
\caption{Sensitivity of the linear $e^{+} e^{-}$ collider
to $ZZ\gamma$ and $Z\gamma\gamma$ couplings at 95\% C.L. for
$\sqrt{s}=1.5$ TeV and $L_{int}=500$ $fb^{-1}$. Only one of the
couplings is assumed to deviate from the SM at a time.\label{tab3}}
\begin{ruledtabular}
\begin{tabular}{ccccc}
$\lambda_{Z}$ &$h_{3}^{Z}$&
$h_{4}^{Z} $ & $h_{3}^{\gamma} $
&$h_{4}^{\gamma}$ \\
\hline
TR+LO &$|4\times 10^{-4}|$& $|3\times 10^{-6}|$&
$|2\times 10^{-4}|$ & $|2\times 10^{-6}|$  \\
(ISR) &$|5\times 10^{-4}|$  & $|4\times 10^{-6}|$ & 
$|3\times 10^{-4}|$ & $|2\times 10^{-6}|$  \\
TR &$|6\times 10^{-3}|$ & - &  $|4\times 10^{-3}|$ &-   \\
(ISR) &$ |7\times 10^{-3}|$ &  -  & $ |4\times 10^{-3}|$ &-  \\
LO &$ |7\times 10^{-5}|$  & $ |5\times 10^{-7}|$ & 
$|4\times 10^{-5}|$  &$ |3\times 10^{-7}|$  \\
(ISR) & $|2\times 10^{-4}|$ & $|2\times 10^{-6}|$  & 
$|1\times 10^{-4}|$  & $|1\times 10^{-6}|$ \\
\end{tabular}
\end{ruledtabular}
\end{table}

\turnpage

\appendix*
\section{Helicity Amplitudes}
There are four Feynman diagrams for the process 
$e^{+}e^{-}\to Z\gamma$ if one includes  $Z\gamma Z^{\star}$, 
$Z\gamma\gamma^{\star}$ vertices.
Helicity amplitudes $M_{1}$ and $M_{2}$ are responsible
for the diagrams concerning  $Z\gamma Z^{\star}$ and 
$Z\gamma\gamma^{\star}$ 
interactions arising from s channel Z or $\gamma$ exchanges. 
$M_{3}$ and $M_{4}$ are  standard model 
contribution of the t and u channel of the process.
The parameters of the helicity amplitudes 
$M(\sigma_{e}, \sigma_{e}^{\prime};
\lambda_{Z},\lambda_{\gamma})$
are helicities of incoming  electron and positron,  
outgoing Z boson  and photon. The values they take are given
by :
\begin{eqnarray}
\sigma_{e}: L, R~~,~~ \sigma_{e}^{\prime}: L, R
 ~~,~~ \lambda_{Z} : +, -, 0 ~~,~~ \lambda_{\gamma} : +, - 
\end{eqnarray}
Here L and R stand for left and right.
Helicity amplitudes we have obtained for each Feynman 
diagram in the c.m. frame of  $e^{+}e^{-}$
can be written as follows:

\begin{eqnarray}
M(\sigma_{e}, \sigma_{e}^{\prime};
\lambda_{Z},\lambda_{\gamma})&&=
M_{1}(\sigma_{e}, \sigma_{e}^{\prime};
\lambda_{Z},\lambda_{\gamma})
+M_{2}(\sigma_{e}, \sigma_{e}^{\prime};
\lambda_{Z},\lambda_{\gamma})
\nonumber \\
&&+M_{3}(\sigma_{e}, \sigma_{e}^{\prime};
\lambda_{Z},\lambda_{\gamma})
+M_{4}(\sigma_{e}, \sigma_{e}^{\prime};
\lambda_{Z},\lambda_{\gamma})
\end{eqnarray}

\begin{eqnarray}
M_{1}(LR;++)&&=-C_{1}^{L} h_{3}^{Z} (M_{Z}^{2}-s) \sin{\theta}
\\
M_{1}(RL;++)&&=-C_{1}^{R} h_{3}^{Z} (M_{Z}^{2}-s) \sin{\theta}
\\
M_{1}(LR;+-)&&=0  \\
M_{1}(RL;+-)&&=0   \\
M_{1}(LR;--)&&=C_{1}^{L} h_{3}^{Z} (M_{Z}^{2}-s) \sin{\theta}
\\
M_{1}(RL;--)&&=-C_{1}^{R} h_{3}^{Z} (M_{Z}^{2}-s) \sin{\theta}
\\
M_{1}(LR;-+)&&=0  \\
M_{1}(RL;-+)&&=0   \\
M_{1}(LR;0+)&&=-\frac{C_{1}^{L}}{\sqrt{2}M_{Z}}
\sqrt{s}(M_{Z}^{2}-s)(1+\cos{\theta})\left[h_{3}^{Z}+
\frac{(M_{Z}^{2}-s)h_{4}^{Z}}{2M_{Z}^{2}}\right] \\
M_{1}(RL;0+)&&=\frac{C_{1}^{R}}{\sqrt{2}M_{Z}}
\sqrt{s}(M_{Z}^{2}-s)(1-\cos{\theta})\left[h_{3}^{Z}+
\frac{(M_{Z}^{2}-s)h_{4}^{Z}}{2M_{Z}^{2}}\right] \\
M_{1}(LR;0-)&&=\frac{C_{1}^{L}}{\sqrt{2}M_{Z}}
\sqrt{s}(M_{Z}^{2}-s)(1-\cos{\theta})\left[h_{3}^{Z}+
\frac{(M_{Z}^{2}-s)h_{4}^{Z}}{2M_{Z}^{2}}\right] \\
M_{1}(RL;0-)&&=-\frac{C_{1}^{R}}{\sqrt{2}M_{Z}}
\sqrt{s}(M_{Z}^{2}-s)(1+\cos{\theta})\left[h_{3}^{Z}+
\frac{(M_{Z}^{2}-s)h_{4}^{Z}}{2M_{Z}^{2}}\right] 
\end{eqnarray}

where
\begin{eqnarray}
C_{1}^{L}=\frac{g_{e}g_{L}}{2M_{Z}^{2}}
~, ~~~
C_{1}^{R}=\frac{g_{e}g_{R}}{2M_{Z}^{2}}
\end{eqnarray}
with
\begin{eqnarray}
g_{L}=\frac{g_{Z}}{2}(C_{V}+C_{A})~, ~~~
g_{R}=\frac{g_{Z}}{2}(C_{V}-C_{A})
\\
C_{V}=2\sin^{2}{\theta_{W}}-\frac{1}{2}~, ~~~C_{A}=-\frac{1}{2}
\\
g_{Z}=\frac{g_{e}}{\sin{\theta_{W}}\cos{\theta_{W}}}
~, ~~~g_{e}^{2}=4\pi\alpha
\end{eqnarray}

\begin{eqnarray}
M_{2}(LR;++)&&=-C_{2} h_{3}^{\gamma} (M_{Z}^{2}-s) \sin{\theta}
\\
M_{2}(RL;++)&&=M_{2}(LR;++)
\\
M_{2}(LR;+-)&&=0 \\
M_{2}(RL;+-)&&=0 \\
M_{2}(LR;-+)&&=0 \\
M_{2}(RL;-+)&&=0 \\
M_{2}(LR;--)&&=C_{2} h_{3}^{\gamma} (M_{Z}^{2}-s) \sin{\theta}
\\
M_{2}(RL;--)&&=M_{2}(LR;--) \\
M_{2}(LR;0+)&&=-\frac{C_{2}}{\sqrt{2}M_{Z}}
\sqrt{s}(M_{Z}^{2}-s)(1+\cos{\theta})\left[h_{3}^{\gamma}+
\frac{(M_{Z}^{2}-s)h_{4}^{\gamma}}{2M_{Z}^{2}}\right]
\\
M_{2}(RL;0+)&&=\frac{C_{2}}{\sqrt{2}M_{Z}}
\sqrt{s}(M_{Z}^{2}-s)(1+\cos{\theta})\left[h_{3}^{\gamma}+
\frac{(M_{Z}^{2}-s)h_{4}^{\gamma}}{2M_{Z}^{2}}\right]
\\
M_{2}(LR;0-)&&=M_{2}(RL;0+) \\
M_{2}(RL;0-)&&=M_{2}(LR;0+) 
\end{eqnarray}

where
\begin{eqnarray}
&&C_{2}=\frac{Q_{e}g_{e}^{2}}
{2M_{Z}^{2}}~,~~~
~,~~~Q_{e}=-1
\end{eqnarray}

\begin{eqnarray}
M_{3}(LR;++)&&=-\frac{C_{3}^{L}s(1-\cos{\theta})\sin{\theta}}
{2M_{e}^{2}+(s-M_{Z}^{2})(1-\cos{\theta})}
\\
M_{3}(RL;++)&&=\frac{C_{3}^{R}[2M_{Z}^{2}-s(1-\cos{\theta})]
\sin{\theta}} {2M_{e}^{2}+(s-M_{Z}^{2})(1-\cos{\theta})}
\\
M_{3}(LR;+-)&&=\frac{C_{3}^{L}s(1-\cos{\theta})\sin{\theta}}
{2M_{e}^{2}+(s-M_{Z}^{2})(1-\cos{\theta})}
\\
M_{3}(RL;+-)&&=-\frac{C_{3}^{R}s(1+\cos{\theta})\sin{\theta}}
{2M_{e}^{2}+(s-M_{Z}^{2})(1-\cos{\theta})}
\\
M_{3}(LR;--)&&=\frac{C_{3}^{L}[2M_{Z}^{2}-s(1-\cos{\theta})]
\sin{\theta}} {2M_{e}^{2}+(s-M_{Z}^{2})(1-\cos{\theta})}
\\
M_{3}(RL;--)&&=-\frac{C_{3}^{R}s(1-\cos{\theta})\sin{\theta}}
{2M_{e}^{2}+(s-M_{Z}^{2})(1-\cos{\theta})}
\\
M_{3}(LR;-+)&&=-\frac{C_{3}^{L}s(1+\cos{\theta})\sin{\theta}}
{2M_{e}^{2}+(s-M_{Z}^{2})(1-\cos{\theta})}
\\
M_{3}(RL;-+)&&=\frac{C_{3}^{R}s(1-\cos{\theta})\sin{\theta}}
{2M_{e}^{2}+(s-M_{Z}^{2})(1-\cos{\theta})}
\\
M_{3}(LR;0+)&&=-\frac{C_{3}^{L}\sqrt{s}(s+M_{Z}^{2})
\sin^{2}{\theta}}
{\sqrt{2}M_{Z}[2M_{e}^{2}+(s-M_{Z}^{2})(1-\cos{\theta})]}
\\
M_{3}(RL;0+)&&=-\frac{C_{3}^{R}\sqrt{2s}
[3M_{Z}^{2}-s+(s+M_{Z}^{2})\cos{\theta}]
\sin^{2}{\frac{\theta}{2}}}
{M_{Z}[2M_{e}^{2}+(s-M_{Z}^{2})(1-\cos{\theta})]}
\\
M_{3}(LR;0-)&&=\frac{C_{3}^{L}\sqrt{2s}
[3M_{Z}^{2}-s+(s+M_{Z}^{2})\cos{\theta}]
\sin^{2}{\frac{\theta}{2}}}
{M_{Z}[2M_{e}^{2}+(s-M_{Z}^{2})(1-\cos{\theta})]}
\\
M_{3}(RL;0-)&&=\frac{C_{3}^{R}\sqrt{s}(s+M_{Z}^{2})
\sin^{2}{\theta}}
{\sqrt{2}M_{Z}[2M_{e}^{2}+(s-M_{Z}^{2})(1-\cos{\theta})]}
\end{eqnarray}

where
\begin{eqnarray}
&&C_{3}^{L}=Q_{e}g_{e} g_{L}
~,~~~C_{3}^{R}=Q_{e}g_{e} g_{R}
\end{eqnarray}

\begin{eqnarray}
M_{4}(LR;++)&&=\frac{C_{4}^{L}[s(1+\cos{\theta})-2M_{Z}^{2}]
\sin{\theta}}{2M_{e}^{2}+(s-M_{Z}^{2})(1+\cos{\theta})}
\\
M_{4}(RL;++)&&=\frac{C_{4}^{R}s(1+\cos{\theta})
\sin{\theta}}{2M_{e}^{2}+(s-M_{Z}^{2})(1+\cos{\theta})}
\\
M_{4}(LR;+-)&&=\frac{C_{4}^{L}s(1-\cos{\theta})
\sin{\theta}}{2M_{e}^{2}+(s-M_{Z}^{2})(1+\cos{\theta})}
\\
M_{4}(RL;+-)&&=-\frac{C_{4}^{R}s(1+\cos{\theta})
\sin{\theta}}{2M_{e}^{2}+(s-M_{Z}^{2})(1+\cos{\theta})}
\\
M_{4}(LR;--)&&=\frac{C_{4}^{L}s(1+\cos{\theta})
\sin{\theta}}{2M_{e}^{2}+(s-M_{Z}^{2})(1+\cos{\theta})}
\\
M_{4}(RL;--)&&=\frac{C_{4}^{R}[s(1+\cos{\theta})-2M_{Z}^{2}]
\sin{\theta}}{2M_{e}^{2}+(s-M_{Z}^{2})(1+\cos{\theta})}
\\
M_{4}(LR;-+)&&=-\frac{C_{4}^{L}s(1+\cos{\theta})
\sin{\theta}}{2M_{e}^{2}+(s-M_{Z}^{2})(1+\cos{\theta})}
\\
M_{4}(RL;-+)&&=\frac{C_{4}^{R}s(1-\cos{\theta})
\sin{\theta}}{2M_{e}^{2}+(s-M_{Z}^{2})(1+\cos{\theta})}
\\
M_{4}(LR;0+)&&=\frac{C_{4}^{L}\sqrt{2s}
[s+(s+M_{Z}^{2})\cos{\theta}-3M_{Z}^{2}]
\cos^{2}{\frac{\theta}{2}}}
{M_{Z}[2M_{e}^{2}+(s-M_{Z}^{2})(1+\cos{\theta})]}
\\
M_{4}(RL;0+)&&=-\frac{C_{4}^{R}\sqrt{s}
(s+M_{Z}^{2})\sin^{2}{\theta}}
{\sqrt{2}M_{Z}[2M_{e}^{2}+(s-M_{Z}^{2})(1+\cos{\theta})]}
\\
M_{4}(LR;0-)&&=\frac{C_{4}^{L}\sqrt{s}
(s+M_{Z}^{2})\sin^{2}{\theta}}
{\sqrt{2}M_{Z}[2M_{e}^{2}+(s-M_{Z}^{2})(1+\cos{\theta})]}
\\
M_{4}(RL;0-)&&=-\frac{C_{4}^{R}\sqrt{2s}
[s+(s+M_{Z}^{2})\cos{\theta}-3M_{Z}^{2}]
\cos^{2}{\frac{\theta}{2}}}
{M_{Z}[2M_{e}^{2}+(s-M_{Z}^{2})(1+\cos{\theta})]}
\end{eqnarray}

where
\begin{eqnarray}
&&C_{4}^{L}=C_{3}^{L}
~,~~~C_{4}^{R}=C_{3}^{R}
\end{eqnarray}

\end{document}